\documentclass{article}
\usepackage{amsmath}                             
\usepackage{amsfonts}
\input psfig.sty
\def\fig{./}
\newtheorem{theorem}{Theorem}

\newtheorem{corollary}{Corollary}

\newtheorem{lemma}{Lemma}

\newtheorem{definition}{Definition}

\newtheorem{remark}{Remark}
\newtheorem{assumption}{Assumption}

\def\Iden{\mbox{$\bf 1\ $}}
\def\n{\noindent}

\def\h {\mathfrak{h}}
\def\m {\mathfrak{m}}
\def\g {\mathfrak{g}}
\def\k {\mathfrak{k}}
\def\s {\mathfrak{s}}
\def\i {\mathfrak{i}}
\def\so {\mathfrak{so}}
\def\su {\mathfrak{su}}

\def\Re {\mathbb{R}}

\def\c {\mathfrak{c}}
\begin{document}

\author{Navin Khaneja,\thanks{Department of Mathematics, Dartmouth College, Hanover, NH 03755, email: navin@hrl.harvard.edu. }\ \ \ Roger Brockett,\thanks{Division of Applied Sciences, Harvard University, Cambridge, MA 02138. This work was funded by the Army grant DAAG 55-97-1-0114, Brown Univ. Army DAAH 04-96-1-0445, and MIT Army DAAL03-92-G-0115.} \ \ \ Steffen J. Glaser\thanks{Institute
of Organic Chemistry and Biochemistry II, Technical University Munich,
85747 Garching, Germany. This work was funded by the Fonds der Chemischen Industrie and the Deutsche Forschungsgemeinschaft under grant Gl 203/1-6.}}

\title{{\bf Time Optimal Control in Spin Systems}}

\maketitle
\begin{center}
{\bf Abstract}
\end{center}
\n In this paper, we study the design of pulse sequences
 for NMR spectroscopy as a problem of time optimal control of the unitary propagator. Radio frequency pulses are used in coherent spectroscopy to implement a unitary transfer of state. Pulse sequences that accomplish a desired transfer
should be as short as possible in order to minimize the effects of
relaxation and
to optimize the sensitivity of the experiments. Here, we give an analytical characterization of such time optimal pulse sequences applicable to coherence transfer experiments in multiple-spin systems. We have adopted a general mathematical formulation, and present many of our results in this setting, mindful of the fact that new structures in optimal pulse design are constantly arising. Moreover, the general proofs are no more difficult than the specific problems of current interest. From a general control theory perspective, the problems we want to study
have the following character. Suppose we are given a controllable right
invariant
system on a compact Lie group, what is the minimum time required to steer the
system from
some initial point to a specified final point? In NMR spectroscopy and quantum
computing, this translates to, what is the minimum time required to produce a
unitary propagator? We also give an analytical characterization of maximum achievable 
transfer in a given time for the two-spin systems. 

\begin{center}
\section{Introduction}\end{center}
Many areas of spectroscopic fields, such as nuclear magnetic
resonance (NMR), electron magnetic resonance and  optical
spectroscopy rely on a limited set of control variables in
order to create desired unitary transformations \cite{Ernst, Science, optics}.
In NMR, unitary transformations are used to
manipulate an ensemble of nuclear spins, e.g. to transfer
coherence between coupled spins in multidimensional
NMR-experiments \cite{Ernst} or to implement quantum-logic
gates in NMR quantum computers \cite{QC}. However, the
design of a sequence of radio-frequency pulses that
generate a desired unitary operator is not trivial \cite{design}. Such a
pulse sequence should be as short as possible in order to
minimize the effects of relaxation or decoherence that are
always present. So far, no general approach was known
to determine the minimum time for the implementation of a
desired unitary transformation \cite{Science}. Here we give an
analytical characterization of such time optimal pulse
sequences related to coherence transfer experiments in
multiple spin systems. We determine, for
example, the best possible in-phase and anti-phase \cite{Science,
transfer} coherence transfer
achievable in a given time. We show 
that the optimal in-phase transfer sequences
improve the transfer efficiency relative to the isotropic mixing sequences \cite{Isotropic} and demonstrate the
optimality of some previously known sequences.

During the last decade the questions of controllability of quantum systems have generated considerable interest \cite{judson,herbruggen}. In particular, coherence or polarization transfer in pulsed coherent spectroscopy has received lot of attention \cite{Science, design}. Algorithms for determining bounds quantifying the maximum possible efficiency of transfer between non-Hermitian operators have been determined \cite{Science}. There is utmost need for design strategies for pulse sequences that can achieve these bounds. From a control theory perspective, this is a constructive controllability problem \cite{khaneja}. At the same time it is desirable that the pulse sequences be as short as possible so as to minimize the relaxation effects. This naturally leads us to the problem of time optimal control, i.e. given that there exist controls that steer the system from a given initial to final state, we would like to determine controls that achieve the task in minimum possible time \cite{herbruggen,khaneja.thesis}. 

In non-relativistic quantum mechanics, the time evolution of a quantum system is defined through the time-dependent Schr{\"o}dinger equation $$\dot{U(t)} = -iH(t)U(t),\ \ U(0)=I,$$ where $H(t)$ and $U(t)$ are the Hamiltonian and the unitary displacement operators, respectively. In this paper, we will only be concerned with finite-dimensional quantum systems. In this case, we can choose a basis and think of $H(t)$ as a Hermitian matrix. We can split the Hamiltonian $$H = H_d + \sum_{i=1}^m v_i(t)H_i,$$ where $H_d$ is the part of Hamiltonian that is internal to the system and we call it the {\it drift} or {\it free Hamiltonian} and $\sum_{i=1}^m v_i(t)H_i$ is the part of Hamiltonian that can be externally changed. It is called the {\it control} or {\it rf Hamiltonian}. The equation for $U(t)$ dictates the evolution of the density matrix according to $$\rho(t) = U(t)\rho(0)U^{\dagger}(t).$$ The problem we are ultimately interested in is to find the minimum time required to transfer the density matrix from the initial state $\rho_0$ to a final state $\rho_F$. Thus, we will be interested in computing the minimum time required to steer the system \begin{equation}\label{eq:unitary}\dot{U} = -i(H_d + \sum_{i=1}^mv_iH_i)\ U, \end{equation}from identity, $U(0)=I$, to a final propagator $U_F$. 

In the following section we establish a framework for studying such problems. For reasons suggested before our approach is more general than the current application requires, but this added generality does not complicate the development. 

\begin{center}
\item \section{Preliminaries}\label{sec:prem}
\end{center}
We will assume that the reader is familiar with the basic facts about Lie groups and homogeneous spaces \cite{nomizu}. Throughout this paper, $G$ will denote a compact semi-simple Lie group and $e$ its identity element (we use I to denote the identity matrix when working with the matrix representation of the group). As is well known there is a naturally defined bi-invariant metric on $G$, given by the Killing form. We denote this bi-invariant metric by $<,>_{G}$. Let $K$ be a compact closed subgroup of $G$. We will denote by $L(G)$ the Lie algebra of right invariant vector fields on $G$ and similarly $L(K)$ the Lie algebra of right invariant vector fields on $K$. There is a one to one correspondence between these vector fields and the tangent spaces $T_e(G)$ and $T_e(K)$, which we denote by $\g$ and $\k$ respectively. Consider the direct sum decomposition $\g = \m + \k$ such that $\m = \k^{\bot}$ with respect to the metric.  

\n To fix ideas, let $G = SU(n)$ and $\g = \su (n)$ be its associated Lie algebra of $n \times n$ traceless skew-Hermitian matrices. Then $<A,B>_{G} = tr(A^{\dagger}B),\ A,B \in \su (n)$ (which is proportional to the Killing metric) represents a bi-invariant metric on $SU(n)$. 

It is well known that the (right) coset space $G/K = \{KU : U \in G \}$ (homogeneous space) admits the structure of a differentiable manifold \cite{nomizu}. Let $\pi: G \rightarrow G/K$ denote the natural projection map. Define $o \in G/K$ by $o = \pi(e)$. Given the decomposition $\g = \m + \k$, there exists a neighborhood of $0 \in \m$ which is mapped homeomorphically onto a neighborhood of the origin $o \in G/K$ by the mapping $\pi \circ \exp |_{\m}$. The tangent space plane $T_{o}(G/K)$ can be then identified with the vector subspace $\m$. The geometry of homogeneous space will play an essential part in determining the shortest possible times for transfers. 

The Lie group $G$ acts on its Lie algebra $\g$ by conjugation $Ad_G : \g \rightarrow \g$ (called the adjoint action). This is defined as follows. Given $U \in G, \ X \in \g$, then $$Ad_U(X) = \frac{d\ U^{-1}\exp(tX)U}{dt}|_{t=0}.$$ Once again to fix ideas if $G = SU(n)$ and $U \in G,\ A \in \su (n)$, then $Ad_U(A) = U^{\dagger}AU$. We use the notation $$Ad_K(X) = \bigcup_{k \in K}Ad_k(X) .$$

If the homogeneous space $G/K$ is a Riemannian symmetric space \cite{wolf}, the Lie algebra decomposition $\g = \m + \k$ (see \cite{Helg} for properties of these orthogonal involutive Lie algebras) satisfies the commutation relation $$[\k,\k] \subset \k ,\ [\m,\k] \subset \m ,\ [\m,\m] \subset \k .$$ If $\h$ is a subalgebra of $\g$ contained in $\m$, then $\h$ is abelian because $[\m, \m] \in \g$. It is well known \cite{wolf} that:

\begin{theorem}\label{th:rank}{\rm If $\h$ and $\h'$ are two maximal abelian subalgebras of $\m$, then 
\begin{enumerate}
\item There exists an element $\xi \in \h$ whose centralizer in $\m$ is just $\h$.

\item There is an element $k \in K$ such that $Ad_k(\h) = \h'$.

\item $\m = \bigcup_{k \in K} Ad_k(\h)$. 
\end{enumerate}}
\end{theorem}

Thus the maximal abelian subalgebras of $\m$ are all $Ad_K$ conjugate and in particular they have the same dimension. The dimension will be called the {\it rank} of the symmetric space $G/K$ and the maximal abelian subalgebras of $\m$ are called the {\it Cartan subalgebras} of the pair $(\g,\k)$. We will see in what follows that the structure of the time optimal control depends on the rank in an important way. We state a useful corollary of the above the theorem \cite{wolf}.

\begin{corollary}\label{cor:rank}{\rm Let $G/K$ be a Riemannian symmetric space. Let $\h$ be a Cartan subalgebra of the pair $(\g,\k)$ and define $A = \exp (\h) \subset G$. Then $G = KAK$.}
\end{corollary}
\n {\bf Proof:} $G = KP$, where $P = \exp(\m)= \exp(\bigcup_{k \in K} Ad_k(\h)) = \bigcup_{k \in K}Ad_k(\exp (\h)) = \bigcup_{k \in K}Ad_k(A) \subset KAK$. Now $G = KKAK = KAK$. \hfill{\bf{Q.E.D}}.  
 
\n Note the space $G/K$ is a union of maximal abelian subgroups $Ad_k(A)$, called {\it maximal tori}. 

\begin{assumption} \label{assum:1} {\rm Let $U \in G$ and let the control system \begin{equation}\label{eq:main}
\dot{U} = [X_d + \sum_{i=1}^{m}v_i X_i ] U , \ U(0)= e
\end{equation}be given. Please note we are working with the matrix representation of the group. We use $\{X_d, X_1, \dots, X_m \}_{LA}$ to denote the Lie algebra generated by $\{X_d, X_1, \dots, X_m \}$. We will assume that $\{X_d, X_1, \dots, X_m \}_{LA} = \g$,  and since $G$ is compact, it follows that the system (\ref{eq:main}) is controllable \cite{jurd}.  Let $\k = \{X_i\}_{LA}$ and $K = \exp \{ X_i\}_{LA}$ be the closed compact group generated by $\{X_i\}$. Given the direct sum decomposition $\g = \m + \k$ where $\m = \k ^{\bot}$ with respect to the bi-invariant metric $<,>_G$, let $X_d \in \m$. We will assume that $Ad_K(\m) \subset \m$, in which case one says the homogeneous space $G/K$ is reductive. All our examples will fall into this category.}
\end{assumption}

\n {\bf Notation:} Let ${\cal C}$ denote the class of all locally bounded measurable functions defined on the interval $[0,\infty)$ and taking value in $\Re^{m}$.  ${\cal C}[0, T]$ denotes their restriction on the interval $[0, T]$. We will assume throughout that in equation (\ref{eq:main}), $v = (v_1, v_2,\dots,v_m)\in {\cal C}$. Given $v \in {\cal C}$, we use $U(t)$ to denote the solution of equation (\ref{eq:main}) such that $U(0)=e$. If, for some time $t \geq 0$, $U(t)= U'$, we say that the control $v$ steers $U$ into $U'$ in $t$ units of time and $U'$ is attainable or reachable from $U$ at time $t$.

\begin{definition}{\bf (Reachable Set):} {\rm The set of all $U' \in G$ attainable from $U_0$ at time $t$ will be denoted by $R(U_0,t)$. Also we use the following notation
\begin{eqnarray*}
{\bf R}(U_0,T) &=& \bigcup_{0\leq t \leq T}R(U_0,t) \\
{\bf R}(U_0) &=& \bigcup_{0\leq t \leq \infty}R(U_0,t).     
\end{eqnarray*}
We will refer to ${\bf R}(U_0)$, as the reachable set of $U_0$.} 
\end{definition}

\begin{remark}{\rm From the right invariance of control systems it follows that $R(U_0,T)=R(e,T)U_0$, ${\bf R}(U_0,T) = {\bf R}(e,T)U_0$, and ${\bf R}(U_0) = {\bf R}(e)U_0$. Note that ${\bf R}(U_0,T)$ need not be a closed set, we use $\overline{{\bf R}(U_0,t)}$ to denote its closure. } \end{remark}

\begin{definition}{\bf (Infimizing Time):} {\rm Given $U_F \in G$, we will define\begin{eqnarray*}
t^{\ast}(U_F) &=& \inf\ \{t \geq 0 |\ U_F \in \overline{{\bf R}(e,t)} \} \\
t^{\ast}(K U_F) &=& \inf\ \{t \geq 0 |\ kU_F \in \overline{{\bf R}(e,t)},\ k \in K\}
\end{eqnarray*}and $t^{\ast}(U)$ is called the {\it infimizing time}.} 
\end{definition}

\n From a mathematical point of view, we may identify two goals in this paper: (1) to characterize $\overline{{\bf R}(e,t)}$ and hence compute $t^{\ast}(U_F)$, the infimizing time for $U_F \in G$, and (2) to characterize the infimizing control sequence $v^{n}$ in (\ref{eq:main}), which in the limit $n \rightarrow \infty$, achieves the transfer time $t^{\ast}(U_F)$ of steering the system (\ref{eq:main}) from identity $e$ to $U_F$. From the physics point of view, these results will help to establish the minimum time required and the optimal controls (the rf pulse sequence in NMR experiments) to achieve desired transfers in a spectroscopy experiment.

\begin{center}
\section{Time Optimal Control}\end{center}

The key observation is the following. In the control system (\ref{eq:main}), if $U_F \in K$ then $t^{\ast}(U_F) = 0$. To see this, note that by letting $v$ in (\ref{eq:main}) be large, we can move on the subgroup $K$ as fast as we wish. In the limit as $v$ approaches infinity, we can come arbitrarily close to any point in $K$ in arbitrarily small time with almost no effect from the term $X_d$. By same reasoning for any $U \in G$, $t^{\ast}(U) = t^{\ast}(kU)$ for $k \in K$. Thus, finding $t^{\ast}(U_F)$ reduces to finding the minimum time to steer the system (\ref{eq:main}) between the cosets $Ke$ and $KU_F$. 

\begin{figure}[t]
\centerline{\psfig{file= \fig/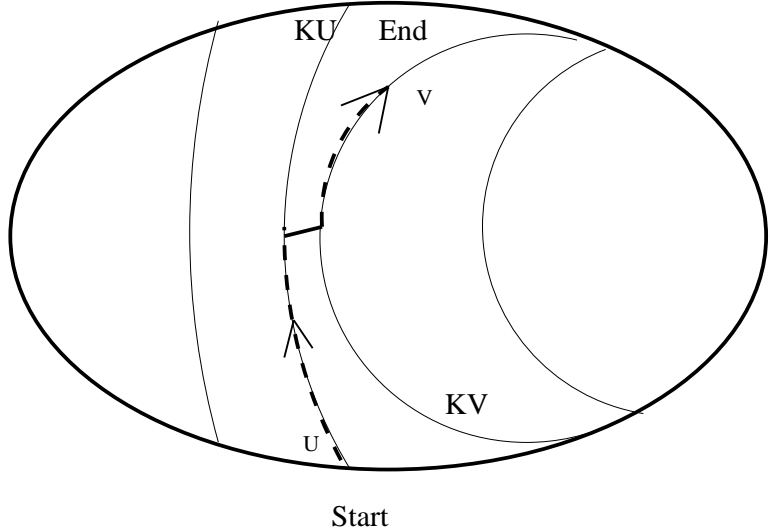 ,width=4in}}
\caption[coset]{The panel 
shows the time optimal path between elements $U$ and $V$ belonging to $G$. The dashed line depicts the fast portion of the path corresponding to movement within the coset $KU$ and, in traditional NMR language, corresponds to the pulse and the solid line corresponds to the slow portion of the curve connecting different cosets and corresponds to evolution of the couplings. 
\label{fig:coset}}
\end{figure}

This is illustrated in the Figure \ref{fig:coset}, where the cosets $KU$ and $KV$ are depicted and the infimizing time path between elements $U$ and $V$ belonging to $G$ is shown. The dashed part of the curve illustrates the fast motion within the coset. The solid part of the curve corresponds to the drift part of the flow ( also known as the evolution of couplings in $NMR$ literature). The minimum time problem then corresponds to finding shortest path between these cosets or, in other words, the shortest path in the space $G/K$. 

\n With this intuitive picture in mind, we now state some lemmas.

\begin{lemma}\label{lem:cont} {\rm Let $U \in G$ and $X : \Re \rightarrow \g$ be a locally bounded measurable function of time. If $X_n(t)$ converges to $X(t)$ in the sense that $$\lim_{n \rightarrow \infty} \int_{0}^T \|X(t) - X_n(t)\| dt = 0 ,$$ then the solution of the differential equation $\dot U = X_n(t)U$ at time $T$ converges to the solution of $\dot U = X(t)U $ at time $T$.} 
\end{lemma}

\n The proof of the above result is a direct consequence of the uniform convergence of the Peano-Baker series. We use this to show

\begin{lemma}\label{lem:mintime}{\rm
For the control system in equation (\ref{eq:main}), $t^{\ast}(U_F) = t^{\ast}(K U_F)$.} 
\end{lemma}
\n {\bf Proof:} We first show that if $k \in K$, then $t^{\ast}(k)=0$. Because $\exp\{X_1, \dots, X_m \}_{LA}= K$, given any $T > 0$ there exists $\bar v \in {\cal C}(T)$, such that the solution of 
$$\dot{U} = [\sum_{i=1}^{m}\bar v_i(t) X_i ] U , \ U(0) = e $$ takes on the value $k$ at time $T$. Now consider the family of control systems $$\dot{U} =  [X_d + \alpha \sum_{i=1}^{m}\bar v_i(\alpha t) X_i ] U , \ U(0) = e.$$ Rescaling time as $\tau = \alpha t$, we obtain

$$\frac{d U}{d \tau} =  [\frac{X_d}{\alpha} + \sum_{i=1}^{m}\bar v_i(\tau) X_i ] U , \ U(0) = e.$$
Observe that, by Lemma \ref{lem:cont}, as $\alpha \rightarrow \infty$, $U(\tau)|_{\tau = T} = k$ or $\lim_{\alpha \rightarrow \infty} U(t)|_{t = \frac{T}{\alpha}} = k$. Therefore $k \in \overline{{\bf R}(e,T)}$, for all $T >0$, implying $t^{\ast}(k)=0$. 

We now prove the general assertion. Let $t^{\ast}(U_F) = T$, we show that if $U_1 = kU_F$ for $k \in K$, then $t^{\ast}(U_1) = T$. Because $t^{\ast}(U_F) = T$, for any $T_1>T$, $U_F \in \overline{R(e,T_1)}$, therefore there exists a family of control laws $v^r[0,T_1]$ such that the corresponding solutions $U^r(t)$ to the equation (\ref{eq:main}) satisfy $U^r(T_1) \rightarrow U_F$. From the first part of the proof, for any $T_2 > T_1$ there exists a control sequence $v^p[T_1, T_2]$ such that the solutions $U^p(t)$ to the family of control systems 
$$\dot U =  [X_d + \sum_{i=1}^{m}v^p_i X_i ] U , \ U(T_1) = U_F $$ satisfies $U^p(t^p) \rightarrow U_1$, for $t^p < T_2$ and $t^p \rightarrow T_1$. Using the continuity of the solution of the differential equation to its initial condition and Lemma \ref{lem:cont}, we conclude that there exists a family of control laws $v^n[0,T_2]$ such that the corresponding solutions $U^n(t)$ to the family of control systems $$\dot U =  [X_d + \sum_{i=1}^{m}v^n_i X_i ] U , \ U(0) = e$$ satisfy $U^n(t^n) \rightarrow U_1$, for $t^n < T_2$. Therefore, $U_1 \in \overline{R(e,T_2)}$. Since $T_2 > T_1$ is arbitrary $t^{\ast}(U_1) \leq T_1$. Because $T_1 > T$ is also arbitrary, we infer that $t^{\ast}(U_1) \leq T$. This shows that $t^{\ast}(U_1) \leq t^{\ast}(U_F)$. Now reverse the roles of $U_F$ and $U_1$ to get the opposite inequality. This proves the claim.  \hfill{\bf{Q.E.D}}.

\n \begin{remark}{\rm The above observation will help us make a bridge between the problem of computing $t^{\ast}(U_F)$ and the problem of computing minimum length paths for a related problem which we now explain.} 
\end{remark}

\begin{definition}{\bf (Adjoint Control System):}
{\rm Let $P \in G$. Associated with the control system (\ref{eq:main}) is the right invariant control system 
\begin{equation}\label{eq:main.1}
\dot{P} = XP, 
\end{equation}where now the control $X$ no longer belongs to the vector space but is restricted to an adjoint orbit i.e., $X \in Ad_{K}(X_d) = \{ k^{-1} X_d k | k \in K \}$. We call such a control system an {\it adjoint control system}}.
\end{definition} 

\n For the control system (\ref{eq:main.1}), we say that $K U_F \in B(U_0,t')$ if there exists a control $X[0,t']$ which steers $P(0)= U_0$ to $P(t') \in KU_F$ in $t'$ units of time. We use the notation $${\bf B}(U_0,T) = \bigcup_{0\leq t \leq T}B(U_0,t).$$ From Lemma \ref{lem:cont}, we see that ${\bf B}(U_0,T)$ is closed. 

\n We use $$L^{\ast}(K U_F) = \inf\ \{t \geq 0 |\ K U_F \in {\bf B}(e,t)\}$$ to denote the minimum time required to steer the system (\ref{eq:main.1}) from identity $e$ to the coset $K U_F$. We call it the {\it minimum coset time}.    

\begin{theorem}\label{th:mintime}{\bf (Equivalence theorem):} {\rm The infimizing time $t^{\ast}(U_F)$ for steering the system $$\dot{U} = [X_d + \sum_{i=1}^{m}v_i X_i ] U $$ from $U(0) = e$ to $U_F$ is the same as the minimum coset time $L^{\ast}(K U_F)$, for steering the adjoint system $$\dot{P} = X P, \ X \in Ad_K(X_d)$$ from $P(0)=e$ to $K U_F$.}
\end{theorem}

\n {\bf Proof:} Let $Q \in K$ satisfy the differential equation \begin{equation}\label{eq:main.2} \dot Q = [ \sum_{i=1}^{m}v_i X_i ] Q , \ Q(0) = e.\end{equation} Let $P \in G$ evolve according to the equation \begin{equation}\label{eq:main.3} \dot P = (Q^{-1} X_d Q)\ P ,\ P(0) = e .\end{equation} Then observe that $$\frac{d (Q\ P)}{dt} = [X_d + \sum_{i=1}^{m}v_i X_i ] (Q P) , \ \ Q(0)P(0) = e, $$ which is the same evolution equation as that of $U$, and since $U(0) = Q(0)P(0)= e$, by the uniqueness theorem for the differential equations, $U(t)= Q(t)P(t)$. Therefore, given a solution $\hat{U}(t)$ of equation (\ref{eq:main}) with the initial condition $\hat{U}(0)$, there exist unique curves $\hat P (t)$ and $\hat Q (t)$, defined through equations (\ref{eq:main.2}) and (\ref{eq:main.3}), satisfying $\hat U(t) = \hat Q(t) \hat P(t)$. Observe that if $\hat U(T) = U_F$ then it follows that $\hat P (T) \in K U_F$. If $U_F \in \overline{{\bf R}(e,T)}$, then there exists a sequence of control laws $v^r[0,T]$ such that the corresponding solutions $U^r(t)$ of (\ref{eq:main}) satisfy $U^r(T) \rightarrow U_F$. Therefore, the solutions $P^r(t)$ of the associated control system (\ref{eq:main.2}) satisfy $\lim_{r \rightarrow \infty}P^r(T) \in KU_F$. Because ${\bf B}(e,T)$ is closed, it follows that  $K U_F \in {\bf B}(e,T)$, which implies that $L^{\ast}(K U_F) \leq t^{\ast}(U_F)$. 

\n To prove the equality observe that if $K U_F \in {\bf B}(e,T)$, then there exists a 
control $\bar X[0,T]$ such that the corresponding solution $\bar P(t)$ to (\ref{eq:main.1}) satisfies $\bar P(T) \in KU_F$. Because $\bar X(t) \in Ad_K(X_d)$, we can express $\bar X(t)$ as ${\bar Q(t)}^{-1}X_d \bar Q(t)$. It is well known \cite{hermes} that we can find a family $v^r(t)$ of control laws such that the corresponding solution $Q^r(t)$ of  
$$\dot{Q^{r}}= [\sum_{i=1}^{m}v^{r}_i X_i]  Q^r,\ Q^r(0) = e $$ 
satisfy $\lim_{r \rightarrow \infty} \int_0^{T}\|\bar Q(t) - Q^{r}(t)\| dt = 0$. Hence, $\lim_{r \rightarrow \infty} \int_0^{T}\|\bar X(t) - (Q^{r}(t))^{-1} X_d Q^{r}(t)\| dt = 0$. Using Lemma \ref{lem:cont}, we claim that the solutions to family of differential equations $$\dot P^{r} = [ (Q^{r})^{-1}(t) X_d Q^{r}(t)] P^{r},\ P^{r}(0) = e $$ satisfies $\lim_{r \rightarrow \infty} P^{r}(T) \in K U_F$. Therefore, $t^{\ast}(K U_F) \leq T$. Since the choice of $T$ was arbitrary, it follows $t^{\ast}(K U_F) \leq L^{\ast}(K U_F)$. Because $t^{\ast}(K U_F) = t^{\ast}(U_F)$, it follows that $t^{\ast}(U_F) \leq L^{\ast}(K U_F)$. Hence the proof. \hfill{\bf{Q.E.D}} \\

\begin{remark}\label{rem:close}{\rm The control system (\ref{eq:main}) evolves on the group $G$ and induces a control system on the coset space $G/K$ through the projection map $\pi$. The adjoint control system (\ref{eq:main.1}) is a representation of this induced control system. Observe that since $\|X\| = 1$ in (\ref{eq:main.1}), we can also define $L^{\ast}(K U_F)$ as the infimizing value of $\int_0^{1}<\dot P,\dot P>^{\frac{1}{2}}dt $ for steering the system $$\dot{P} = \gamma XP ,\ \gamma > 0$$
from $P(0)= e$ to $P(1) \in KU_F$.} 
\end{remark}

\n We will now compute $t^{\ast}(U_F)$ using the properties of the set $Ad_K(X_d)$. Based on the qualitative nature of time optimal trajectories of the system (\ref{eq:main}), we make the following classification. 

\begin{enumerate}

\item {\bf Riemannian Symmetric Case} In addition to Assumption \ref{assum:1}, if we have the restriction $[\m, \m] \subset \k $, then we are in the Riemannian symmetric case as described in the section \ref{sec:prem}. We can further classify this case based on the rank of the symmetric space $G/K$.

\begin{itemize}

\item {\bf Pulse-drift-pulse sequence}(characteristic of single-spin systems) In this case, the rank of the symmetric space $G/K$ is one. Roughly speaking the trajectories of the infimizing control sequence $v^r$ (which in the limit $r \rightarrow \infty$, achieves the transfer time $t^{\ast}(U_F)$) converge to an impulse (which resembles an impulse of appropriate shape), followed by evolution under drift (for time $t^{\ast}(U_F)$) and a final impulse. \\

\item {\bf Chained Pulse-drift-pulse sequence}( characteristic of two-spin system)In this case, the rank of the symmetric space $G/K$ is more than one. The trajectories corresponding to an infimizing control sequence $v^r$ in (\ref{eq:main.1}) converge to a chain of `` impulse drift impulse'' pattern. The infimizing time $t^{\ast}(U_F)$ is the time spent when the system just evolves under drift. \\
\end{itemize}

\item {\bf Chatter sequence} In this case, $G/K$ is no more a Riemannian symmetric case, i.e. $[\m, \m] \not\subset \k $. This is a characteristic of more that two-spin systems.

\end{enumerate}

In this paper we will confine ourselves to the Riemannian symmetric case. The non-symmetric case will be treated in detail in a forthcoming paper.

\n {\bf Pulse-drift-pulse sequence}

\n We begin with the first case where the rank of the symmetric space $G/K$ is one. It follows from Theorem \ref{th:rank} that $\m = \bigcup_{\alpha \geq 0}Ad_K(\alpha X_d)$. In this case, computing $t^{\ast}(U_F)$ reduces to finding the geodesic distance on the homogeneous space $G/K$. Given the bi-invariant metric $<,>_G$ on $G$, there is a corresponding left invariant metric $<,>_n$, on the homogeneous space $G/K$ arising from the restriction of $<,>_G$ to $\m$. Let $L_n(\gamma)$ represent the length of a curve $\gamma \in G/K$ under the standard induced metric. There is a one-to-one correspondence between the curves $\{ \gamma(t) \in G/K | \gamma(0)= 0,\ \gamma(1)= \pi(U_F) ,\ L_n(\gamma[0,1]) = T\}$ and the trajectories of system (\ref{eq:main.1}) satisfying $\{P(0)= e , P(T) \in KU_F \}$. Therefore, $L^{\ast}(K U)$ is the Riemannian distance between $\pi(e)$ and $\pi(U)$ under the standard metric $<,>_n$. This is computed in the following theorem, which characterizes geodesics on the homogeneous space $G/K$
 under the standard metric \cite{nomizu}.

\begin{theorem}\label{th2:mintime}{\rm 
Let $G$ be a compact Lie group with a bi-invariant metric $<,>$, and $K$ be a closed subgroup. Let $\g$ and $\k$ denote their Lie algebras with the direct sum decomposition $\g = \m + \k,\ \m = \k^{\bot}$. Consider the right invariant control system $$\dot{U} = [X_d + \sum_{i=1}^{m}v_i X_i ] U ,\ U \in G, \ U(0)= e$$ where $v_i \in \Re$, $X_d \in \m$, and $\{X_i \}_{LA} = \k$. Suppose $G/K$ is a Riemannian symmetric space of rank one, then $ t^{\ast}(U_F)$ is the smallest value of $\alpha > 0$ such that we can solve $U_F = Q_1 \exp(\alpha X_d) Q_2 $ with $Q_1, Q_2 \in K$.}
\end{theorem}

\n {\bf Proof:} By the equivalence theorem $t^{\ast}(U_F) = L^{\ast}(K U_F)$, where  $L^{\ast}(K U_F)$ is the minimum time for steering the system $$\dot{P} = X P, \ X \in Ad_K(X_d)$$ from $P(0)=e$ to $K U_F$. Because $G/K$ is a Riemannian symmetric space of rank one, $L^{\ast}(K U_F)$ is the Riemannian distance between $o$ and $\pi(U)$ under the standard metric $<,>_n$. From \cite{nomizu}, the geodesics under the metric $<,>_n$ originating from $o$ take the form $\pi(\exp(\tau X))$ for $X \in \m$. If $U_F =  Q_1 \exp(t X_d) Q_2$ for $Q_1,\ Q_2 \in K$, then $\pi(U_F) = \pi(\exp(t\ Q_2^{-1}X_d Q_2))$. To see this note that $U_F =  (Q_1Q_2)Q_2^{-1}\exp(t X_d) Q_2 =  (Q_1Q_2)\exp(t\ Q_2^{-1}X_d Q_2)$. Thus, from the above assertion, $\pi(\exp(\tau \ Q_2^{-1}X_d Q_2))$ is the unique geodesic connecting $o$ to $\pi(U_F)$ and has the length $L = t$. Hence the proof. \hfill{\bf{Q.E.D}} \\  

\begin{remark}{\rm Roughly speaking, the time optimal trajectory (obtained as a limit of the infimizing sequence) for the system (\ref{eq:main}), which steers the system form $U(0)=e$ to $U_F = Q_1 \exp(\alpha X_d) Q_2$, takes the form $e \rightarrow Q_2 \rightarrow \exp(\alpha X_d) Q_2 \rightarrow Q_1\exp(\alpha X_d) Q_2$, where the first and last step of this chain takes no time, and the time is required for the drift process(second step)}.
\end{remark}

\noindent We now use illustrate these ideas through some examples. \\

\noindent \begin{corollary} \label{cor:min0}{\rm Let $U \in G = SU(2)$, and let $ I_x = \frac{1}{2} \left[ \begin{array}{cc}
0 & 1 \\
1 & 0 
\end{array}\right]$ , $I_z = \frac{1}{2}\left[  
\begin{array}{cc}
1 & 0 \\
0 & -1 \end{array}\right]
$ represent the Pauli spin matrices. Given the unitary evolution of the single-spin system 
$$\dot{U} = -i[I_z + v I_x] U,$$ where the control $v \in \Re$. Let $U_x = \exp(-iI_x t)$ represent the one-parameter subgroup generated by $I_x$. Given $U_F \in SU(2)$, there exists a unique $\beta \in [0,\ 2\pi]$ such that $U_F = U_1 \exp[-i \beta I_x] U_2$, where $U_1, U_2 \in U_x$. The infimizing time $t^{\ast}(U_F) = \beta$.} 
\end{corollary} 

\n {\bf Proof:} First note that the Lie algebra $\g = \su (2)$ has the decomposition $\m = \{iI_y, iI_z \}$, $\k = \{iI_x\}$, and $Ad_{U_x}(I_z)= \m$. Observe from corollary \ref{cor:rank} that any $U_F \in SU(2)$ has a representation $U_F = Q_1 \exp[-i\alpha I_z] Q_2$, where $Q_1, Q_2 \in U_x$. Because $\exp[-itI_z]$ is periodic with period $4\pi$, the $\beta$ with the smallest absolute value for which $U_F = U_1 \exp[-i \beta I_z] U_2$ holds, belongs to the interval $[-2\pi, 2\pi]$. Because $-I_z \in Ad_{U_x}(I_z)$, we can restrict $\beta$ to the interval $[0, 2\pi]$. The proof then follows directly from the Theorem \ref{th2:mintime}. \hfill{\bf{Q.E.D.}}\\

\noindent \begin{corollary} \label{cor:min1}{\rm Let $\Theta \in G = SO(3)$, and let $ \Omega_x = \left[ \begin{array}{ccc}
0 & -1 & 0 \\
1 & 0 & 0 \\
0 & 0 & 0
\end{array}\right]$ , $\Omega_z = \left[  
\begin{array}{ccc}
0 & 0 & 0  \\
0 & 0  & -1 \\
0 & 1 & 0
\end{array}\right]
$ represent the generators of rotation around $x-$ and $z-$axis. Consider the control system 
$$\dot{\Theta} = [\Omega_z + v \Omega_x] \Theta, $$ where the control $v \in \Re$. Let $\Theta_x = \exp(\Omega_x t)$ represent the one-parameter subgroup generated by $\Omega_x$. Given $\Theta_F \in SO(3)$, there exists a unique $\beta \in [0,\ \pi]$ such that $\Theta_F = Q_1 \exp[\beta \Omega_x] Q_2$, where $Q_1, Q_2 \in \Theta_x$. The infimizing time $t^{\ast}(\Theta_F) = \beta$.} 
\end{corollary} 

\n {\bf Proof:} First note that the Lie algebra $\g = \so (3)$ has the decomposition $\m = \{\Omega_y, \Omega_z \}$, $\k = \{\Omega_x\}$, and $Ad_{\Theta_x}(\Omega_z)= \m$. Observe that any $\Theta_f \in SO(3)$ has a representation $\Theta_f = Q_1 \exp[\alpha \Omega_z] Q_2$, where $Q_1, Q_2 \in \Theta_x$. Because $\exp[t\Omega_z]$ is periodic with period $2\pi$, the proof is on the same lines as Corollary \ref{cor:min0}. \hfill{\bf{Q.E.D.}}\\

\n We now generalize the example to the case where $G = SO(n)$, the group of $n \times n$ orthogonal matrices. The Lie algebra is $\g = \so (n)$, the set of $n \times n$ skew-symmetric matrices. The bi-invariant metric on $G$ is $<\Omega, \Omega> = tr(\Omega^{T}\Omega)$. Consider the following decomposition of $\g $. Let $\m$  consists of skew-symmetric matrices which are zero except the first row and column and $\k$  consists of skew symmetric matrices which are zero in the first row and column. Observe that $\k$ generates the subgroup $SO(n-1)$. Then we have 

\begin{corollary}\label{cor:min2}{\rm Let $\Theta \in G = SO(n)$ and let the control system 
 $$\dot{\Theta} = [\Omega_d + \sum_{i=1}^{m}v_i \Omega_i ] \Theta , \ \Theta(0) = I$$ be given, where $\Omega_d \in \m$, $v_i \in \Re$, $\Omega_i \in \k$ and $\Omega_d \in \m$, such that $\exp[t\Omega_d]$ has period $2\pi$. Suppose that $K = \exp \{\Omega_i\}_{LA} = SO(n-1)$. Given $\Theta_F \in SO(n)$, there exists a unique $\beta \in [0,\ \pi]$ such that $\Theta_F = \Theta_1 \exp[\beta \Omega_x] \Theta_2$, where $\Theta_1, \Theta_2 \in \Theta_x$. The infimizing time $t^{\ast}(\Theta_F) = \beta$.} 
\end{corollary}

\n {\bf Proof:} Observe that $Ad_{K}(\Omega_d) = \m$ and hence the proof is on the same lines as Corollary \ref{cor:min0}. \hfill{\bf{Q.E.D.}}\\

\n {\bf Chained Pulse-drift-pulse sequence}

\n Let us now consider the second case in our classification scheme. We now analyze the case when the rank of the Riemannian symmetric space $G/K$ is greater than one. 

\n {\bf Definition:} {\bf (Weyl Orbit)} Given the decomposition $\g = \m + \k$, let $\h \subset \m$ represent the maximal abelian subalgebra containing $X_d$. We use the notation $\Delta_{X_d} = \h \bigcap Ad_K(X_d)$ to denote the maximal commuting set contained in the adjoint orbit of $X_d$. The set $\Delta_{X_d}$ is called the Weyl orbit of $X_d$. We use $\c(X_d) = \{ \sum_{i=1}^{n}\beta_i X_i | \beta_i \geq 0, \sum \beta_i = 1, X_i \in \Delta_{X_d}\}$, to denote the convex hull of the Weyl orbit of $X_d$, with vertices given by the elements of the Weyl orbit of $X_d$.

We compute the infimizing time for the system (\ref{eq:main}), in the following Theorem (\ref{th2:mintime2}), which is a generalization of the rank one case. 

\begin{remark} {\rm Recall from corollary (\ref{cor:rank}) that, if $A = \exp(\h)$, where $\h$ is the maximal abelian subalgebra contained in $\m$, then $G = KAK$.
Therefore given any $U_F \in G$, we can express $U_F = Q_1 \exp(Z) Q_2 = Q_1Q_2 \exp(Ad_{Q_2}(Z))$, where 
$ Q_1, \ Q_2 \in K$ and $Z \in \h$. Suppose $Z = \sum_{i=1}^{n}\beta_i X_i,\ \ \beta_i \geq 0, X_i \in \Delta_{X_d}$. By choosing $X(t)$ to be $Ad_{Q_2}(X_i)$ for $\beta_i$ units of time we can steer the adjoint control system $\dot{P} = X(t)P$ from the identity to the coset $KU_F = K \exp(Ad_{Q_2}(Z))$. The claim of the following theorem is that indeed the fastest way to get to the coset $KU_F$ is to flow on the maximal torus, $Ad_{Q_2}(A),\ Q_2 \in K$, containing the coset $KU_F$.}
\end{remark}

\n We now state a convexity theorem due to Kostant \cite{kostant}, which is the main idea behind the following theorem.

\begin{theorem}{\bf(Kostant's Convexity Theorem):} {\rm Given the direct sum decomposition $\g = \m + \k$, let $\h \subset \m$ represent a maximal abelian subalgebra containing $X_d \in \m$. Let $\Gamma : \m \rightarrow \h$, be the orthogonal projection of $\m$ onto $\h$. Then $\Gamma : Ad_K(X_d) = \c (X_d)$, where $\c (X_d)$ is the convex hull of the Weyl orbit of $X_d$ as defined above.}
\end{theorem}

\begin{theorem}\label{th2:mintime2}{\bf (Time Optimal Tori Theorem:)} {\rm 
Let $G$ be a compact matrix Lie group and $K$ be a closed subgroup with $\g$ and $\k$ their Lie algebras, respectively such that $G/K$ is a Riemannian symmetric space. Let the direct sum decomposition $\g = \m + \k$, such that $\m = \k^{\bot}$, be given. Consider the right invariant control system $$\dot{U} = [X_d + \sum_{i=1}^{m}v_i X_i ] U ,\ U \in G, \ U(0)= e,$$ where $v_i \in \Re$, $X_d \in \m$, $\{X_i \}_{LA} = \k$. Then any $U_F = Q_1\ \exp(\alpha Y)Q_2$, where $\alpha >0$, $ Q_1, \ Q_2 \in K$, and $Y \in \c(X_d)$, belongs to the closure of the reachable set. The infimizing time $t^{\ast}(U_F)$ is the smallest value of $\alpha > 0$, such that we can solve $$U_F= Q_1\ \exp(\alpha Y )Q_2 ,$$ where $ Q_1, \ Q_2 \in K$ and $Y$ belongs to the convex hull $\c(X_d)$ .}
\end{theorem}

We sketch here the outline of a proof. A rigorous proof from a control theoretic viewpoint will be presented elsewhere. 

Once again, we will compute $t^{\ast}(U_F)$ by finding the minimum time it takes to steer the system $\dot{P} = XP$ from $X(0)=e$ to the coset $KU_F$. Following the notation of the corollary \ref{cor:rank}, let $A = \exp (\h)$ denote the maximal torus contained inside $G/K$. From corollary \ref{cor:rank}, we know that $G/K = Ad_K(A)$. Thus it suffices to prove the theorem for $U_F \in A$, therefore we will show that if $U_F \in A$ then $T$ is the smallest value of $\sum_{i=1}^m \alpha_i$, $\alpha_i \geq 0$ such that $U_F = \exp (\sum_{i=1}^m \alpha_i X_i)$, where $X_i \in \Delta_{X_d}$, then $t^{\ast}(U_F) = T$. It is immediate that in the adjoint control system $\dot P = XP$, by letting $X$ to be $X_i$ for $\alpha_i$ units of time we can reach $U_F$ in $T$ units of time. All we need to show is that we can reach the coset $KU_F$ no sooner. Let $\bar{P}(t)$ be the shortest (or time optimal) trajectory of the adjoint control system $\dot P = XP$ that steers $P(0) = e$ to the coset $KU_F$. Let $\pi_A : G/K \rightarrow A$, denote the projection on the maximal subgroup, whereby for $A_1 \in A$ ,  $\pi_A : k^{-1}A_1 k \rightarrow A_1$ (note that the projection is only unique modulo a Weyl group action, hence an explicit choice is required to make the projection unique). Let $ a(t) \in A$ be a continuous path obtained from the projection of $\bar P(t)$, onto $A$. Then observe the projection $\pi_A$ induces the map $\pi_{A_\ast} : \dot{\bar{P}}(t) \rightarrow \dot{a}(t)$. Then the evolution of the curve $a(t)$ can be expressed as $\dot{a}(t) = \Omega\ a(t)$, where $\Omega = \Gamma(Ad_{\tilde k}(X_d))$, for some $\tilde k \in K$ (recall $\Gamma: \m \rightarrow \h$ is the orthogonal projection onto $\h$). Now using Kostant's convexity theorem, we have  $\Omega \in \c(X_d)$. Therefore we can write $\dot{a}(t) = (\sum_{i=1}^{m}\beta_i X_i)\ a(t)$, where $X_i \in \Delta_{X_d}$ and $\sum_{i=1}^{m}\beta_i = 1$ for $\beta_i \geq 0$. This implies that if $T = \sum_{i=1}^m\alpha_i$ is the smallest value for which $U_F \in A$ satisfies $U_F = \exp (\sum_{i=1}^m\alpha_i X_i)$ for $\alpha_i \geq 0$, then the path $a(t)$ will atleast takes $T$ units of time to reach $U_F$. 

\begin{remark}{\rm The theorem characterizes ${\bf B}(e,t)$, the reachable set for the adjoint system. This is given by 
$$K {\bf B}(e,t) = K\exp(\alpha  Y)K , \ 0 \leq \alpha \leq t$$ where $Y$ belongs to the convex hull $\c (X_d)$.}  
\end{remark}  

\begin{center}
\item \section{Spin Algebra}\label{sec:algebra}
\end{center} 
The Lie Group which we will be most interested in is $SU(2^{n})$, the special unitary group describing the evolution of $n$ coupled spins $\frac{1}{2}$. Its Lie algebra $\su (2^n)$ is a $4^n -1$ dimensional space of traceless $n \times n$ skew-Hermitian matrices. The orthonormal basis which we will use for this space is expressed as tensor products of Pauli spin matrices \cite{sorenson}

\begin{eqnarray*}
\label{pauli}
I_x &=& \frac{1}{2}\left(  
\begin{array}{cc}
 0 & 1\\
1 & 0 
\end{array}
\right)
\\  
I_y &=& \frac{1}{2}\left(  
\begin{array}{cc}
 0 & -i\\
i & 0 
\end{array}
\right)\\
I_z &=& \frac{1}{2}\left(  
\begin{array}{cc}
 1 & 0\\
0 & -1 
\end{array}
\right)
\end{eqnarray*}
The matrices ($I_x$, $I_y$, $I_z$) are the generators
for rotation in the two dimensional Hilbert space and  basis for the Lie algebra of
traceless skew-Hermitian matrices $\su (2)$. They obey the well known commutation relations  
$$
 [I_x \ I_y] = i I_z \; \; ; \; \; [I_y \ I_z] = i I_x \; \; ; \;
\;   [I_z \ I_x] = i I_y. $$

Then the basis for $\su (2^n)$ takes the form $\{iB_s \}$ where
\begin{equation}
B_s = 2^{q-1}\prod_{k=1}^{n}(I_{k\alpha})^{a_{ks}},
\end{equation}$\alpha = x, y, or \ z$ and  
\begin{equation}
\label{tensor}
 I_{k\alpha} = \Iden \otimes \cdots \otimes I_{\alpha} \otimes \Iden,
\end{equation}
where $I_{\alpha}$ the Pauli matrix appears in the above expression only at the $k^{th}$ 
position, and $\Iden $ the two dimensional identity matrix, appears everywhere 
except at the $k^{th}$ position. $a_{ks}$ is $1$ for $q$ of the indices and $0$ for the 
remaining. Note that $q \geq 1$ as $q = 0$ corresponds to the identity matrix and is not a part of the algebra. As an example for $n=2$ the basis for $\su (4)$ takes the form 
\begin{eqnarray*}
q = 1 &\ & I_{1x}, I_{1y}, I_{1z}, I_{2x}, I_{2y}, I_{2z}\\
q = 2 &\ & 2I_{1x}I_{2x}, 2I_{1x}I_{2y}, 2I_{1x}I_{2z} \\ 
&\ & 2I_{1y}I_{2x},2I_{1y}I_{2y}, 2I_{1y}I_{2z} \\
&\ & 2I_{1z}I_{2x},2I_{1z}I_{2y}, 2I_{1z}I_{2z}. 
\end{eqnarray*}

It is important to note that these operators are only normalized for $n=2$ as

$$tr(B_r B_s) = \delta_{rs}2^{n-2}.$$

To fix ideas, lets compute one of these operators explicitly for $n=2$ 

\[I_{1z} = \frac{1}{2}\left[  
\begin{array}{cc}
1 & 0 \\
0 & -1
\end{array}
\right] \otimes \left[  
\begin{array}{cc}
1 & 0 \\
0 & 1
\end{array}
\right]   
\]  
which takes the form 
\[I_{1z} = \frac{1}{2}\left[  
\begin{array}{cccc}
1& 0 & 0 & 0\\
 0& 1 & 0 & 0\\
0& 0 & -1 & 0 \\
0 & 0 & 0 & -1
\end{array}
\right].   
\]

We will often refer to the algebra of $\su(2^n)$ as the {\it spin algebra}.  
   
\begin{center}
\item \section{Optimal Transfer in Two-Spin Systems}
\end{center}
\n In this section, we will apply our general results on the time optimal control for the specific case of a heteronuclear two-spin system. In particular, we consider the following important heteronuclear two-spin system discussed in detail in \cite{Science}. By going to a rotating frame, the free evolution part of the Hamiltonian has been reduced to just a scalar coupling evolution. The system then takes the following form. 

Let $U \in SU(4)$, which evolves as
\begin{equation} \label{eq:het.two} \dot{U} = -i (\ H_d + \sum_{i=1}^{4}u_i H_i \ )U , \end{equation} where \begin{eqnarray*}H_d &=& 2\pi J I_{z} S_z \\
H_1 &=& 2\pi I_x \\
\ H_2 &=& 2\pi I_{y} \\
\ H_3 &=& 2\pi S_x \\
H_4 &=& 2\pi S_{y},
\end{eqnarray*}where $I_x$, $I_y$ and $I_z$ represent operators for the first spin and have the same meaning as $I_{1x}$, $I_{1y}$ and $I_{1z}$, respectively, as explained in previous section \ref{sec:algebra}. Similarly $S_x$, $S_y$, and $S_z$ represent operators for the second spin and have the same meaning as $I_{2x}$, $I_{2y}$ and $I_{2z}$. The symbol $J$ represents the strength of the scalar coupling between $I$ and $S$. Observe that the subgroup $K$ generated by $\{H_i \}_{i=1}^{4}$ is $SU(2)\times SU(2)$. 

\n We first compute the infimizing time for steering the system (\ref{eq:het.two}).

\begin{theorem}\label{ex:op2spin.transfer}{\rm
For the heteronuclear spin system, described by the equation (\ref{eq:het.two}), the infimizing time $t^{\ast}(U_F)$ is the smallest value of $\sum_{i=1}^3 \alpha_i$, $\alpha_i > 0$, such that we can solve $$U_F= Q_1\exp(-i2 \pi J (\alpha_1 I_{x}S_x + \alpha_2 I_yS_y + \alpha_3 I_{z}S_{z}))Q_2,$$ where $Q_1$ and $Q_2$ belong to $K$.}  
\end{theorem}

\n {\bf Proof:} Consider the direct sum decomposition $\g = \m + \k$, where $\m = span\{I_{\alpha} S_{\beta} \}$, $\k = span\{I_{\alpha},\ S_{\beta}\}$, and $(\alpha, \beta) \in (x,y,z)$. Then observe $[\m, \m] \in \k$, $[\m, \k] \in \m$, and $[\k, \k] \in \k$. Furthermore, observe that $\Delta_{I_{z}S_z} = \{\pm I_{z}S_z, \pm I_xS_x, \pm I_{y}S_{y} \}$, therefore for $\alpha \geq 0$, we have $Ad_K(\alpha\ \c (X_d)))= \m$. Thus the above example satisfies all the conditions of the theorem \ref{th2:mintime2}. Hence the proof. \hfill{\bf{Q.E.D}} 

\n Now we address the question of maximum possible achievable transfer in some given time $T$. For this purpose we define the transfer efficiency.

\begin{definition}{\bf (Transfer Efficiency):} {\rm Given the evolution of the density matrix $\rho(t) = U(t)\rho(0)U^{\dagger}(t)$, where $$\dot{U} = -i (\ H_d + \sum_{i=1}^{m}u_i H_i \ )U ,\ U(0)=I, $$ define the {\it transfer efficiency} $\eta(t)$ to some given target operator $F$ as $$\eta(t) = \|tr(F^{\dagger}U(t)\rho(0)U^{\dagger}(t))\|.$$} 
\end{definition}

\begin{remark}{ In the formula for the transfer efficiency, we always assume that the starting operator $\rho(0)$ and the final operator $F$ are both  normalized to have norm one (i.e. $tr(F^{\dagger}F) = 1$).}
\end{remark}

\n We will now look at the in-phase and anti-phase transfers in the two-spin system, whose evolution is given by equation (\ref{eq:het.two}). We give here expressions for maximum transfer efficiencies. We first prove some lemmas, which will be required in computing transfer efficiencies.

\begin{lemma}\label{lem:le1}{\rm let $p = \left[  
\begin{array}{c}
1 \\
-i \\
0 
\end{array}
\right]$ and let $\Sigma$ be a real diagonal matrix $$\Sigma = \left[  
\begin{array}{ccc}
a_1 & 0 & 0\\
0 & a_2 & 0 \\
0 & 0 & a_3
\end{array}
\right]. $$ If $a_i \geq a_j  \geq a_k \geq 0$, where $\{i, j, k\} \in \{1, 2, 3 \}$ and let $U, V \in SO(3)$, then the maximum value of $\|p^{\dagger}U \Sigma V p\|$ is $a_i + a_j$.}
\end{lemma}

\n {\bf Proof:} Let $$\Lambda = \left[  
\begin{array}{ccc}
\sqrt{a_1} & 0 & 0\\
0 & \sqrt{a_2} & 0 \\
0 & 0 & \sqrt{a_3}
\end{array}
\right]. $$ By definition $\Sigma = \Lambda^{\dagger}\Lambda$.  Using Cauchy Schwartz inequality $\|p^{\dagger}U \Sigma V p\| \leq \|\Lambda V p\| \ \|\Lambda U p\|$. Observe, the maximum value of $\|\Lambda V p\|$ is $\sqrt{a_i + a_j}$. Therefore $\|p^{\dagger}U \Sigma V p\| \leq a_i + a_j$. Clearly for appropriate choice of $U$ and $V$, this upper bound is achieved (For example, in case $a_1 \geq a_2 \geq a_3$, the bound is achieved for $U$ and $V$ identity). Hence the result follows. \hfill{\bf{Q.E.D}}.

\begin{lemma}\label{lem:le2}{\rm Consider the function $f(\alpha_1, \alpha_2,\alpha_3) = \sin (J \pi \alpha_1) \sin (J \pi \alpha_2) + \sin (J \pi \alpha_1) \sin (J \pi \alpha_3)$. If $\alpha_1, \alpha_2, \alpha_3 \geq 0$ and $\alpha_1 + \alpha_2 + \alpha_3 = T$, where $T \leq \frac{3}{2J}$, then the maximum value 
of $f(\alpha_1, \alpha_2, \alpha_3)$ is $2 sin(J\pi a)sin(J \pi b)$, where $a + 2b = t$ and $\tan(J\pi a) = 2 \tan(J \pi b)$.}  
\end{lemma}

\n {\bf Proof:}\ Let $$H(\alpha_1,\alpha_2, \alpha_3,\lambda) = \sin (J \pi \alpha_1) \sin (J \pi \alpha_2) + \sin (J \pi \alpha_1) \sin (J \pi \alpha_3) + \lambda(\alpha_1 + \alpha_2 + \alpha_3 - T).$$ The necessary condition for optimality gives
$\frac{\partial H}{\partial \alpha_1} = 0$ , $\frac{\partial H}{\partial \alpha_2}=0$, $\frac{\partial H}{\partial \alpha_3}= 0$, which imply respectively that 
\begin{eqnarray}\label{lem2.eq1}
\pi J (\cos (J \pi \alpha_1) \sin (J \pi \alpha_2) &+& \cos (J \pi \alpha_1) \sin (J \pi \alpha_3)) + \lambda = 0 \\ \label{lem2.eq2}
\pi J (\sin (J \pi \alpha_1) \cos (J \pi \alpha_2)) &+& \lambda = 0 \\ \label{lem2.eq3}
\pi J (\sin (J \pi \alpha_1) \cos (J \pi \alpha_3)) &+& \lambda = 0  
\end{eqnarray}
From equation (\ref{lem2.eq2}) and (\ref{lem2.eq3}), we obtain that either $\sin (J \pi \alpha_1) = 0$ or $ \cos (J \pi \alpha_2)) = \cos (J \pi \alpha_3))$. The first condition does not give a maxima as it makes $f$ identically zero. The second condition implies \begin{equation}\label{lem2:eq4} J \pi \alpha_2 = 2m\pi + J \pi \alpha_3. \end{equation} Since $\alpha_2, \alpha_3 \geq 0$ and $\alpha_2 + \alpha_3 \leq T \leq \frac{3}{2J}$, condition (\ref{lem2:eq4}) is only satisfied for $m=0$. Therefore, $\alpha_1 = \alpha_2$. Now substituting this in (\ref{lem2.eq1}) and using the equations (\ref{lem2.eq1}) and (\ref{lem2.eq2}), we get the desired result \hfill{\bf{Q.E.D}}.

\begin{theorem}\label{th:max.inphase}{\bf (Maximum in-phase transfer)} {\rm
Consider the evolution for the heteronuclear IS spin system as defined by Equation (\ref{eq:het.two}). Let $\rho(0)= \frac{S_x - iS_{y}}{\sqrt{2}}$ and $F = \frac{I_x - iI_{y}}{\sqrt{2}}$. For $t \leq \frac{3}{2J}$, the maximum achievable transfer $$\eta^{\ast}(t) = sin(J\pi a)sin(J \pi b),$$ where $a + 2b = t$ and $\tan(J\pi a) = 2 \tan(J \pi b)$. For $t \geq \frac{3}{2J}$ the maximum achievable transfer is one.}  
\end{theorem}

\n {\bf Proof:} Let $$\Lambda(\alpha_1, \alpha_2, \alpha_3) = \exp(-i2\pi J(\alpha_1 I_{x}S_x + \alpha_2 I_yS_y + \alpha_3 I_{z}S_{z})).$$ From now on we will simply write $\Lambda(\alpha_1, \alpha_2, \alpha_3)$ as $\Lambda$. From Theorem \ref{ex:op2spin.transfer}, any unitary propagator $U_F$ belonging to the set
$${\bf R}(e,t) = \{ Q_1\Lambda Q_2 |\ Q_1, \ Q_2 \in K\ \ \alpha_i > 0,\ \sum_{i=1}^{3} \alpha_i \leq t \},$$ can be produced by appropriate pulse sequence in (\ref{eq:het.two}). Therefore we will maximize $\|tr(F^{\dagger}U(t)\rho(0)U^{\dagger}(t))\|$, for $U(t) \in {\bf R}(e,t)$. Let $I = \exp\{iI_x,iI_{y},iI_{z}\}$ and $S = \exp\{iS_x,iS_{y},iS_z\}$. By definition, $K = S \times I$. In the expression 
$$\eta(t) = \|tr(Q_1^{\dagger}F^{\dagger}Q_1\Lambda Q_2 \rho(0) Q_2^{\dagger}\Lambda^{\dagger})\| ,$$ $\rho(0)$ commutes with $I$ and $F$ commutes with $S$, therefore it suffices to restrict $Q_1$ and $Q_2$ to $I$ and $S$, respectively. 

\n Let $\s$ denote the subspace spanned by the orthonormal basis $\{S_x,S_y,S_z\}$ and $\i$ denote the subspace spanned by the orthonormal basis $\{I_x,I_y,I_z\}$. We represent the starting operator $\rho(0) = \frac{1}{\sqrt 2}(S_x - iS_y) $ as a column vector $p = \frac{1}{\sqrt 2} [1\ -i\ 0]^{T}$ in $\s$.
The action $\rho(0) \rightarrow Q_2 \rho(0) Q_2^{\dagger}$ can then be represented as $p \rightarrow V p$ where $V$ is a orthogonal matrix.

\n Let $P_{I}$ denote the projection on the subspace $\i$. A simple computation yields that 
\begin{eqnarray*}
P_{I}(\Lambda S_x \Lambda^{\dagger}) &=& \sin (J \pi \alpha_2) \sin (J \pi \alpha_3) I_x \\
P_{I}(\Lambda S_y \Lambda^{\dagger}) &=& \sin (J \pi \alpha_1) \sin (J \pi \alpha_3) I_y \\
P_{I}(\Lambda S_z \Lambda^{\dagger}) &=& \sin (J \pi \alpha_2) \sin (J \pi \alpha_3) I_z.
\end{eqnarray*}

We denote the target operator $F = \frac{1}{\sqrt 2}(I_x - iI_y)$ as a column vector $\frac{1}{\sqrt 2} [1\ -i\ 0]^{T}$ in $\i$. The action $\rho(0) \rightarrow P_I ( \Lambda Q_2 \rho(0) Q_2^{\dagger} \Lambda^{\dagger})$ can be written as 
$p \rightarrow \Sigma V p$, where 
$$  
\Sigma = \left[  
\begin{array}{ccc}
\sin (J \pi \alpha_2) \sin (J \pi \alpha_3) & 0 & 0\\
0 & \sin (J \pi \alpha_1) \sin (J \pi \alpha_3) & 0 \\
0 & 0 & \sin (J \pi \alpha_1) \sin (J \pi \alpha_2) 
\end{array}
\right], $$

Therefore we can rewrite $\eta(t) = \|tr(Q_1^{\dagger}F^{\dagger}Q_1\Lambda Q_2 \rho(0) Q_2^{\dagger}\Lambda^{\dagger})\| $ as $\eta(t) = \|p^{\dagger}U \Sigma V p\|,$ where $U$ and $V$ are real orthogonal matrices. Using the result of Lemma 
(\ref{lem:le1}), we get that for $\sin (J \pi \alpha_1) \geq \sin (J \pi \alpha_2) \geq \sin (J \pi \alpha_3) \geq 0$, the maximum value of $\eta (t)$ is $$\frac{\sin(J \pi \alpha_1) \sin(J \pi \alpha_2) + \sin(J \pi \alpha_1) \sin(J \pi \alpha_3)}{2}.$$ 
Now we maximize the above expression with respect to $\alpha_1, \alpha_2, \alpha_3$ as worked out in Lemma \ref{lem:le2} to get the above result. 

Now we prove the last part of the theorem. Note for $t = \frac{3}{2J}$, the maximum achievable transfer is one. Because $\rho(0)$ and $F$ are normalized, this is the maximum possible transfer between these operators. If $t > \frac{3}{2J}$, say $t = T + \frac{3}{2J}$, we can always arrange matters so that $U(T) = e$ (\ by creating a propagator $U(T/2) = exp(-i2\pi J(\frac{T}{2}I_{z}S_z))$ and then creating its inverse $exp(i2\pi J(\frac{T}{2}I_{z}S_z))$ from $T/2$ to $T$\ ). In the remaining $\frac{3}{2J}$ units of time, we can produce the optimal propagator.

\hfill{\bf{Q.E.D}}

\begin{figure}[t]
\centerline{\psfig{file= \fig/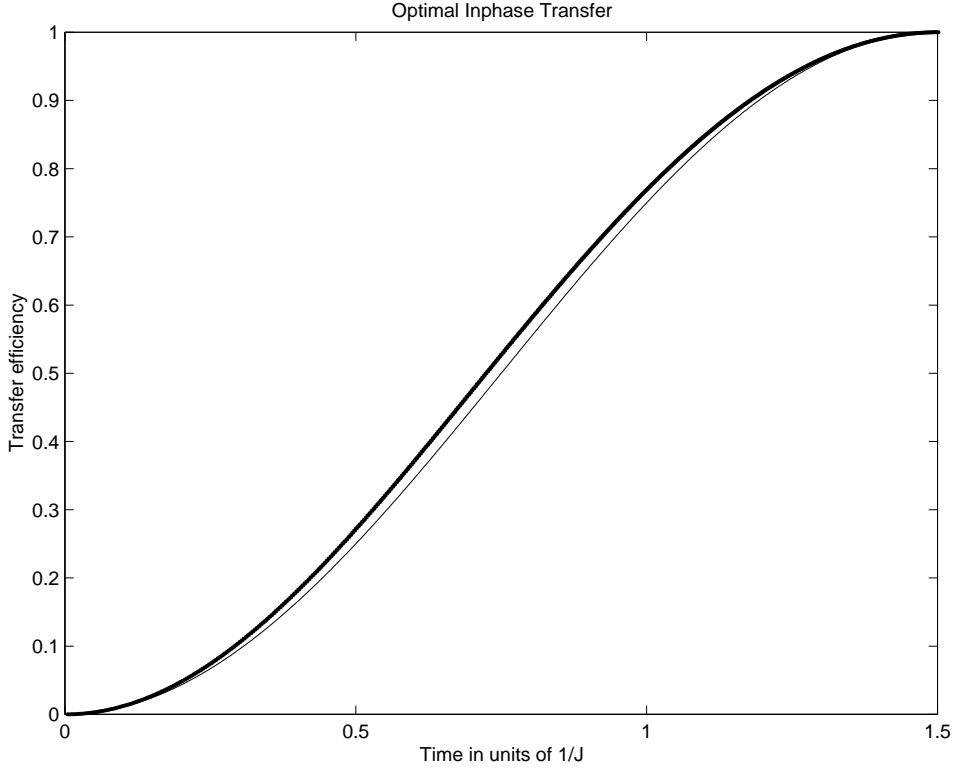 ,width=5in}}
\caption[coset]{The panel 
shows the comparison between the best achievable transfer (bold curve) and the transfer achieved using the isotropic mixing Hamiltonian for the in-phase transfer in 2 spin case. On X axis is plotted time in units of $1/J$.  
\label{fig:2bound}}
\end{figure}

The optimal transfer curve is plotted in comparison with the transfer achieved using the isotropic mixing Hamiltonian in the Figure \ref{fig:2bound}. The unitary propagator $U(t)$ in the isotropic mixing Hamiltonian case takes the form 
$$U(t) = \exp(-i\frac{2\pi Jt}{3}(I_{z}S_z + I_xS_x + I_{y}S_{y})).$$

\n For small mixing times the transfer amplitude achieved by the
optimal experiment is up to 12.5 \% larger than the transfer achieved by
isotropic mixing. This is a previously unknown result that will find immediate practical applications in NMR spectroscopy.

\begin{theorem}{\bf (Maximum anti-phase transfer)}{\rm
 Consider the evolution for the heteronuclear IS spin system as defined by equation (\ref{eq:het.two}). Let $\rho(0)= \sqrt{2}I_zS^{-}= \sqrt{2}I_z(S_x - iS_y)$ and $F = I^{-} = \frac{I_x - iI_y}{\sqrt{2}}$. Then, for $t \leq 1/J$, the  maximum achievable transfer $\eta^{\ast}(t)$ is  $$ \|tr(F^{\dagger}U(t)\rho(0)U^{\dagger}(t))\| = sin(J\pi t/2).$$ For $t \geq \frac{1}{J}$, the maximum achievable transfer is one. }  
\end{theorem}

\n {\bf Proof:} Let $$\Lambda = \exp(-i2\pi J(\alpha_1 I_xS_x + \alpha_2 I_yS_y + \alpha_3 I_zS_z)).$$ From theorem \ref{ex:op2spin.transfer}
$$U(t) \in \{ Q_1\Lambda Q_2 |\ Q_1, \ Q_2 \in K\ \alpha_i > 0,\ \sum_{i=1}^{3} \alpha_i \leq t\}.$$ Let $S = \exp\{iS_x,iS_y,iS_z\}$ and $I = \exp\{iI_x,iI_y,iI_z\}$. By definition $K = S \times I$. In the expression for $$ \eta = \|tr(F^{\dagger}Q_1\Lambda Q_2 \rho(0) Q_2^{\dagger}\Lambda Q_1^{\dagger})\|, $$ 
let $Q_2 = Q_{2I} \times Q_{2S}$, where $Q_{2I} \in I$ and $Q_{2S} \in S$. 
Let the optimal ${Q_{2}}^{\ast} = {Q_{2I}}^{\ast} \times {Q_{2S}}^{\ast}$ be such that $${Q_{2I}}^{\ast}\rho(0){{Q_{2I}}^{\ast}}^{\dagger} = {Q_{2I}}^{\ast}I_zS^{-}{{Q_{2I}}^{\ast}}^{\dagger}= a_z I_zS^{-} + a_y I_yS^{-} + a_x I_xS^{-}, $$ where $a_x^2 + a_y^2 + a_z^2 = 1$.
Denote
\begin{eqnarray*}
\eta_z &=& \|tr(F^{\dagger}(Q_1 \Lambda Q_{2S})^\ast (\sqrt{2} I_zS^{-})(Q_{2S}^{\dagger}\Lambda^{\dagger}Q_1^{\dagger})^{\ast}\ )\| \\ 
\eta_y &=& \|tr(F^{\dagger}(Q_1 \Lambda Q_{2S})^\ast (\sqrt{2} I_yS^{-})(Q_{2S}^{\dagger}\Lambda^{\dagger}Q_1^{\dagger})^{\ast}\ )\| \\
\eta_x &=& \|tr(F^{\dagger}(Q_1 \Lambda Q_{2S})^\ast (\sqrt{2} I_xS^{-})(Q_{2S}^{\dagger}\Lambda^{\dagger}Q_1^{\dagger})^{\ast}\ )\|. \\
\end{eqnarray*}

Then observe that $$ \eta(t) \leq a_z \eta_z + a_y \eta_y + a_x \eta_x .$$

We first compute the maximum of $\eta_z$. Let $P_{I}$ denote the projection on the subspace generated by $\{I_x,I_y,I_z\}$, then a simple computation yields 
\begin{eqnarray*}
P_{I}(\Lambda I_zS_x \Lambda^{\dagger}) &=& \frac{1}{2} \sin (J \pi \alpha_1)I_y \\
P_{I}(\Lambda I_zS_y \Lambda^{\dagger}) &=& \frac{1}{2} \sin (J \pi \alpha_2)I_x \\
P_{I}(\Lambda I_zS_z \Lambda^{\dagger}) &=& 0.
\end{eqnarray*}
Since $\{I_zS_x,I_zS_y,I_zS_z\}$ forms an orthogonal pair, we can rewrite
$$\|tr(F^{\dagger}Q_1\Lambda Q_{2S} (\sqrt 2 I_zS^{-})(Q_{2S}^{\dagger}\Lambda^{\dagger}Q_1^{\dagger})\ )\|$$ as $$\eta(t) = \|p^{\dagger}U \Sigma V p\| ,$$ where $p = [1\ -i\  0]^{T}$, 
$$  
\Sigma = \left[  
\begin{array}{ccc}
\frac{\sin (J \pi \alpha_2)}{2} & 0 & 0\\
0 & \frac{\sin (J \pi \alpha_1)}{2} & 0 \\
0 & 0 & 0 
\end{array}
\right],$$ and $U$ and $V$ are real orthogonal matrices. From Lemma 
\ref{lem:le1}, it follows that the maximum value of $\eta_z$ is $$\frac{\sin (J \pi \alpha_2)}{2} + \frac{\sin (J \pi \alpha_1)}{2}.$$ We can compute the maximum of the above expression under the constraint $\alpha_1 + \alpha_2 = t \leq 1/(J)$. The maximum value of the above expression is obtained for $\alpha_1 = \alpha_2$. The maximum value is $\sin (J \pi t/2)$ for $t \leq 1/J$. Similarly, the maximum value of $\eta_x$ and $\eta_y$ is as above. Since $a_x^2 + a_y^2 + a_z^2 = 1$, we get the desired result. 

\n The final proposition of the theorem has the same proof as in Theorem \ref{ex:op2spin.transfer} \hfill{\bf{Q.E.D}}.

\begin{figure}[t]
\centerline{\psfig{file= \fig/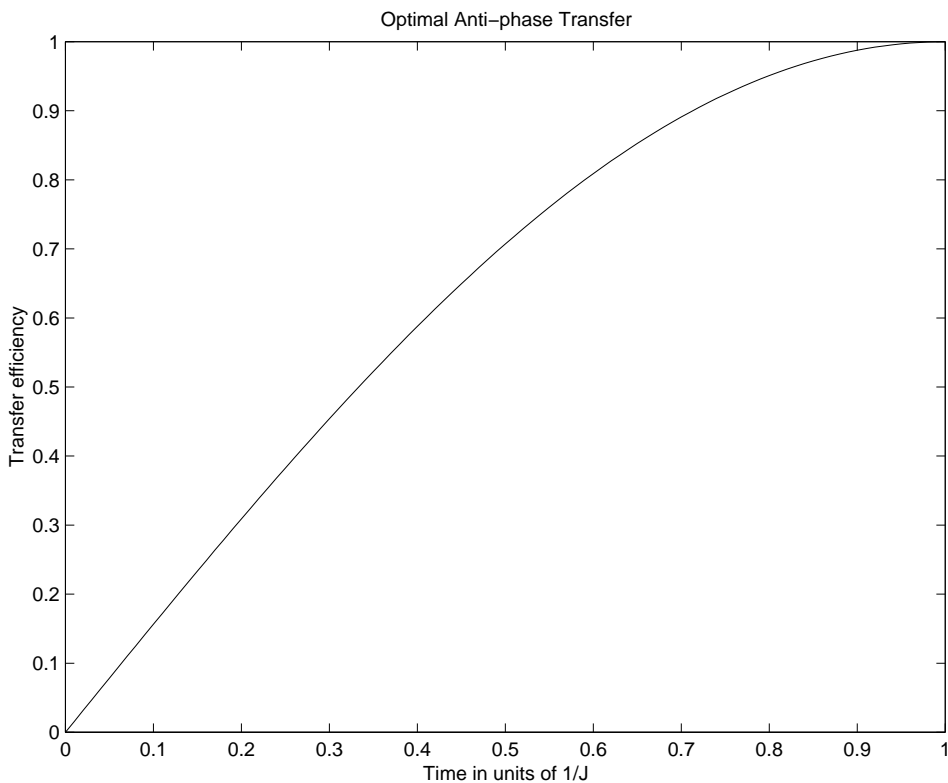 ,width=5in}}
\caption[coset]{The panel 
shows the best achievable transfer as a function of time measured in units of $1/J$ for the anti-phase transfer in 2 spin case. 
\label{fig:2bound.anti}}
\end{figure}

The optimal transfer curve for the anti-phase transfer plotted as a function of time measured in units of $1/J$ is shown in the Figure \ref{fig:2bound.anti}.This clearly shows that the transfer efficiency achieved using the known mixing sequence \cite{transfer} is optimal.

\begin{center}
\section{Conclusion}\end{center}

In this paper, we presented a mathematical formulation of the problem of finding shortest pulse sequences in coherent spectroscopy. We showed how the problem of computing minimum time to produce a unitary propagator can be reduced to finding shortest length paths on certain coset spaces. A remarkable feature of time optimal control laws is that they are singular, i.e. the control is zero most of the time, with impulses in-between. We explicitly computed the shortest transfer times and maximum achievable transfer in a given time for the case of heteronuclear two-spin transfers. In a forthcoming paper, we plan to extend these
results to higher spin systems.

\end{document}